# Planetary boundaries of consumption growth: Declining social discount rates

Victor E. Gluzberg[(1)] and Yuri A. Katz[(2)]

PTC [(1)]

5 Nahum Hat Street, PO Box 15019, 3190500 Haifa, Israel

victor@ptc.com

S&P Global Market Intelligence[(2)]

55 Water Str., New York, NY 10040, USA

yuri.katz@spglobal.com

We introduce the logistic model of consumption growth, which captures a negative feedback loop preventing an unlimited growth of consumption due to finite biophysical resources of our planet. This simple dynamic model allows for derivation of the expression describing the declining long-term tail of a social discount curve. The latter plays a critical role in, e.g., climate finance with benefits on current investments deferred to centuries from now. The growth rate of consumption is irregularly evolving in time, which makes estimation of an *expected* term-structure of consumption growth and associated social discount rates a challenging task. Nonetheless, observations show that the problem at hand is perturbative with the small parameter being the product of an average strength of fluctuations in the growth rate and its autocorrelation time. This fact permits utilization of the cumulant expansion method to derive remarkably simple expressions for the term-structure of expected consumption growth and associated discount rates in the bounded economy. Comparison with empirical data shows that the dynamic effect related to the planetary resource constrains could become a dominant mechanism responsible for a declining long-term tail of a social discount curve at the time horizon estimated here as about100 years from now (the lower boundary).  The derived results can help to shape a more realistic long-term social discounting policy. Furthermore, with the obvious redefinition of the key parameters of the model, obtained results are directly applicable for description of expected long-term population growth in stochastic environments.

I.   **Introduction**

How should governments discount the present costs of long-term public projects, especially those that affect future generations? The core principle of finance holds that the present cost of the future benefit is worth *less* the longer investor should wait to receive it. In the normative social context, governments acting on behalf of society are *exponentially* discounting the present value of future socioeconomic benefits, because future generations are expected to *consume exponentially more*. It is implicitly assumed that an unbounded linear growth of log-consumption will last indefinitely into the future. When exponential discounting is applied to cost-benefit analysis of long-term environmental projects the present value of payoffs deferred to the distant future can become negligible. The actual value critically depends on the choice of a discount rate. But what should the long-term social discount rate be and should it be constant over time? Despite obvious importance of the rigorous social discounting policy, currently there is neither a consensus about the valuation methodology nor even a clear definition of the time horizon that might be considered as the 'distant future'. This problem is especially acute in the context of the climate finance with benefits on current investments to reduce greenhouse gas emissions from human activities deferred to centuries from now [1,2].

The classic Ramsey optimization of the social welfare functional leads to a risk-free social discount rates which are proportional to rates of real (corrected for inflation) per capita consumption (RPCC) growth [2,3,4]. Following insights of Quetelet and Verhulst on population growth [5], we can infer that in the long-run a negative feedback loop – the "Iron Law of Verhulst" - should prevent unlimited growth of consumption and reduce its rate. Thus, eventually a rate of RPCC growth should become a decreasing function of time, leading to a declining term-structure of long-term social discount rates. To quantify this basic insight, we introduce the nonlinear (logistic) model of consumption growth and derive the simple expression describing a social discount curve. The latter reduces to the renowned Ramsey formula in the special case of unbounded economy.

The growth rate of RPCC depends on several macroeconomic factors and is irregularly evolving in time which makes estimation of an *expected* term-structure of consumption growth and associated social discount rates a challenging task. In this paper, we demonstrate that the conventional stochastic models of a consumption growth process are not applicable for description of empirical data [6]. The underlining complex stochastic process cannot be assumed to be either Gaussian, or i.i.d., or even stationary. On the other hand, observations of a consumption growth process show that the problem at hand has a small parameter - the product of an average strength of fluctuations in the growth rate and its autocorrelation time. This fact permits utilization of the powerful formalism of cumulant expansion [7 8 9 10] to derive remarkably simple, stochastic-model-independent expressions for the term-structure of expected consumption growth and associated discount rates in the economy with constrains imposed by natural boundaries of Earth. Comparison with empirical data [6] shows that the dynamic effect related to the planetary resource constrains could become a dominant mechanism responsible for a declining long-term tail of a social discount curve at the time horizon estimated here as about100 years from now (the lower boundary).

We now mention some related works. In the hypothetical limitless economy the only mechanism of a declining long-term tail of a discount curve is related to uncertainty of a social planner in future RPCC growth rates, see Refs.[1 2] and references therein. This 'precautionary' effect is very general. Qualitatively, it drives a government facing uncertainty in an impact of climate changes on a future consumption growth to invest more in mitigation projects today, thereby decreasing the discount rate. Formally, it reflects the Jensen inequality, which due to convexity of the exponential discount factor with negative argument, requires riskless discount rates to be less than their historical average. Very different models accounting for stochastic behavior of the consumption growth process lead to the qualitatively same outcome: the long-term asymptotic of risk-free discount rates is lower than estimates based on the simple average of a consumption growth rate [11 12 13 14 15 16 17 18 19 20 21 22]. Similar conclusion regarding the value of long-term risk-free interest rates has been made within the investment-based approach to social discounting, see Refs. [23 24 25 26] and references therein.

The method of cumulant expansion has been applied by Martin [27] to the problem of long-term discounting in the boundless economy. Account of higher cumulants of the log-consumption growth process allows extending the standard model beyond the commonly used assumption of normality of this process. Note, however, that Martin has considered only the i.i.d. model of the log-consumption growth process. Since all cumulants of any i.i.d. process – Gaussian or not – are linearly growing with time, this study leads to a time-invariant discount rate. Conversely, in the auto-regressive model accounting for memory effects in a consumption growth process, the method of cumulant expansion yields a declining term-structure of a discount curve [28].

We begin Section II with a brief review of the Ramsey optimization framework and the consumption-based approach to discounting. Then, we introduce the logistic model of a consumption growth and derive the simple formula describing a social discount curve. Finally, we consider an irregular evolution of a consumption growth rate and apply the Kubo cumulant expansion method [8] to derive general expressions for the expected term-structure of RPCC growth and associated discount rates. In Section III, we analyze the particular limiting cases and demonstrate how the derived general formulas reduce to the known results. In Section IV, we provide a detailed comparison of our theoretical results with empirical data. In particular, we explore statistical properties of the observed U.S. consumption growth process and find an empirical support for our key assumption: the product of a variance of a growth rate of RPCC and its squared autocorrelation time is very small. We conclude in Section V.

## II. Consumption growth and social discounting in the bounded economy
### a. Ramsey framework in the intergenerational social context

The consumption-based capital asset pricing model connects discount rates with a marginal utility growth of a representative economic agent [4]. The utility $u_t = u(c_t)$ is an increasing and concave function of an agent's consumption, $c_t$, which is randomly changing in time, $\{c_t : t \geq 0\}$ In the *intergenerational* social context of the climate finance, a government plays the role of an agent acting on behalf of society. The latter should reduce the present level of consumption in favor of benefits deferred to a distant future. The connection between social discount rates and marginal utility emerges from quite simple and general considerations, which we expose here for the readers' convenience.

In order to evaluate the social desirability of a project at the present date 0 a government should maximize the Ramsey's wellness functional

$$W \equiv E\left[\int_0^T e^{-\delta t} u(c_t) dt\right] \quad (2.1)$$

Here $\delta$ denotes the rate of preference for the present of a social decision-maker, $T$ is the project's life span, and the expectation operator $E$ is conditional on the information set available at $t = 0$.

Note that it is implicitly assumed here that the current generation is considering benefits obtained by the next ones as its own. In the context of long-term environmental projects one could assume $T = \infty$, provided convergence of the integral in Eq.(2.1). Let us assume that investing of a small (relative to GDP) amount $\xi$ into a project at the present date *guarantees* some proportional net benefits (dividends) in the future, which implies change in the consumption function $c_t$:

$$c'_t = c_t - p_0 \xi \delta(t) + \xi D_t, \qquad (2.2)$$

where $p_0$ is the price of the unit of an investment, $D_t$ denotes a continuous stream of dividends per the same unit, and $\delta(t)$ is the Dirac delta function. In other words, it is assumed here that in order to make an investment into a project, a government must cut RPCC for a short period of time and compensate the cost of this investment in the future (as illustrated by Fig. 1).

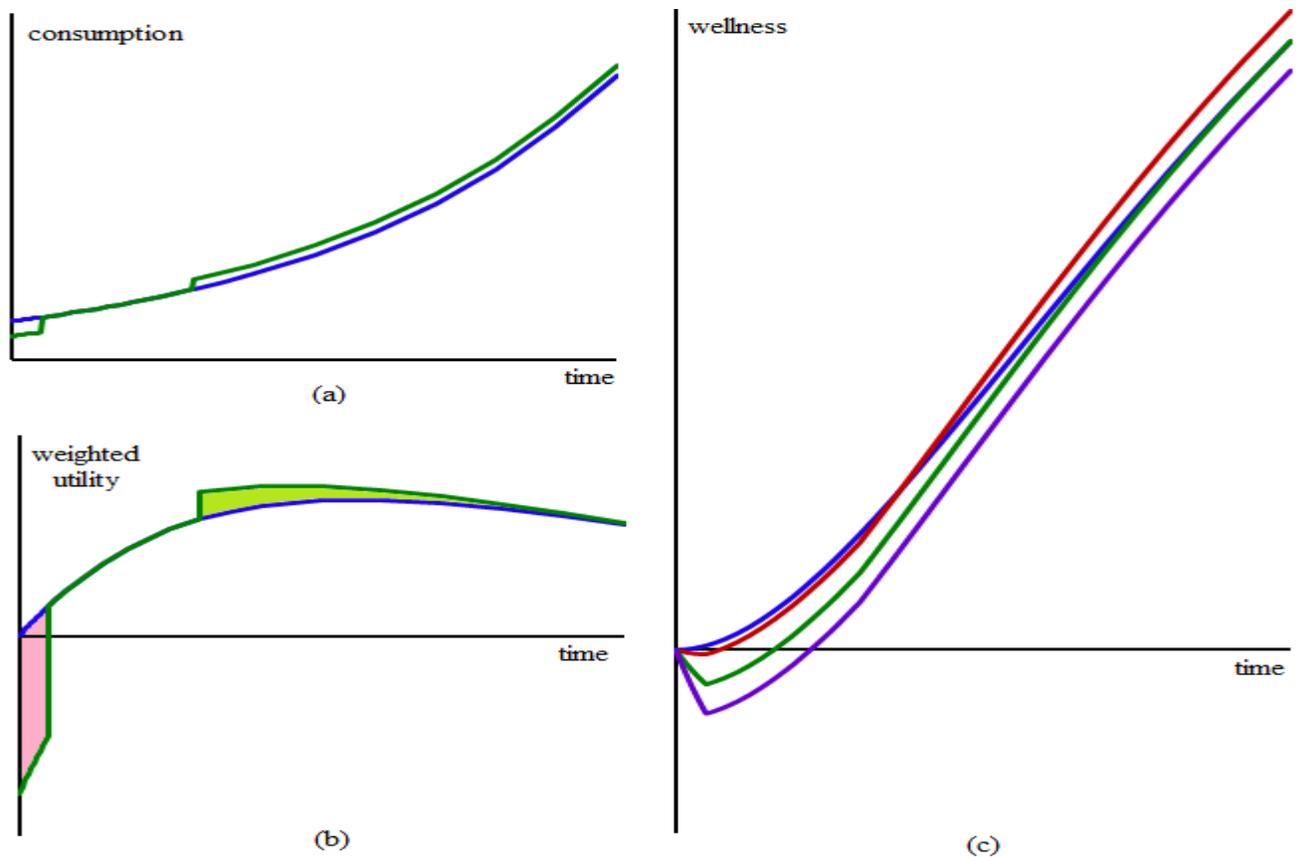

Figure 1. (Color online). To derivation of the discount factor: (a) disturbance of the consumption rate by a small investment yielding dividends in the future, originally anticipated

($c_t$) – blue line, disturbed ($c'_t$) –green line; (b) weighted utility ($e^{-\delta t} u(c'_t)$) - pink and green areas depict initial negative and future positive impact of the investment on the total wellness, respectively; (c) wellness functional as function of time (i.e. an integral of the weighted utility from the origin to the current time) for different investment amounts and prices, no investment – blue line, the fair price - green line (this line converges with the no-investment line at the end of the project lifespan), under- and over-estimated prices - red and purple lines, respectively.

The fair price of an investment is determined by the condition of exact compensation, i.e., by the equilibrium: $\left.\frac{\partial W'}{\partial \xi}\right|_{\xi=0} = 0$, where $W'$ is the wellness functional for the *changed* consumption function $c'_t$:

$$W' \equiv E\int_0^\infty e^{-\delta t'} u(c_{t'} - p_0 \xi \delta(t') + \xi D_{t'}) dt' \tag{2.3}$$

Now, to perform the accurate limiting transition, let us substitute here

$$\delta(t) \equiv \lim_{\tau \to 0} \frac{1}{\tau} \Delta\left(\frac{t}{\tau}\right),$$

where $\Delta(\cdot)$ is an arbitrary regular delta function approximation (for example, like one depicted in Fig. 1(a)), and take *first* the limit with $\xi \to 0$, *then* with $\tau \to 0$. This way, for the initial consumption $c_0$, which is known with certainly, we obtain

$$\left.\frac{\partial W'}{\partial \xi}\right|_{\xi=0} = -p_0 u'(c_0) + E\int_0^\infty e^{-\delta t'} u'(c_{t'}) D_{t'} dt' \tag{2.4}$$

Hence, the equilibrium condition gives

$$p_0 = E\int_0^\infty e^{-\delta t'} \frac{u'(c_t)}{u'(c_0)} D_{t'} dt' \tag{2.5}$$

In the particular case, when all benefits are obtained in a *single sure* (risk-free) payoff at some future moment of time $t > 0$, the function $D_t$ can be modeled by $D \cdot \delta(t' - t)$, leading to the conventional "net present value" formula [2]

$$e^{-y_t t} \equiv \frac{p_0}{D} = e^{-\delta t} E\left[\frac{u'(c_t)}{u'(c_0)}\right] \tag{2.6}$$

This expression determines the risk-free social discount curve as follows

$$y_t = \delta - t^{-1} \ln E \frac{u'(c_t)}{u'(c_0)} \tag{2.7}$$

In the context of the climate finance, it reflects how a government values socioeconomic benefits deferred to the distant future relative to the present cost of investments related to reduction of greenhouse gas emissions from human activities.

Substitution of the commonly used in the capital investment literature "power" utility function with the constant degree of concavity, $\gamma \geq 0$,

$$u(c_t) = [1 - (c_t/c_0)^{1-\gamma}]/(\gamma - 1) \tag{2.8}$$

into Eq.(2.7) yields, see Refs.[2,4]

$$y_t = \delta - t^{-1} \ln E (c_t/c_0)^{-\gamma} \tag{2.9}$$

Note that since future generations are expected to consume more, in the context of long-term social projects, $\gamma$ is interpreted as the intergenerational inequality aversion [2]. The growth rate of $c_t$ depends on a number of macroeconomic factors and is irregularly evolving in time which makes estimation of expected term-structure of social discount rates a challenging task.

### b. Logistic model of consumption growth

A consumption growth process is fundamentally constrained by finite environmental resources of our planet. Therefore, a realistic model of long-term consumption growth should take into account that the relative growth rate of RPCC, $c_t^{-1} dc_t/dt$, must eventually decrease with time. To quantify this effect, we introduce the logistic dynamic model of a consumption growth process. In this model, the evolution of the relative growth rate of RPCC is governed by the Verhulst equation

$$c_t^{-1} \frac{dc_t}{dt} = g_t \left(1 - c_t/C\right) \tag{2.10}$$

Here $g_t$ denotes a generally time-varying and random intrinsic rate of consumption growth, $C$ is the carrying capacity of RPCC. The factor $(1 - c_t/C)$ in the RHS of Eq.(2.10) is specific to the logistic model. It represents a negative feedback loop, which prevents unlimited exponential growth of consumption in a finite world. For an arbitrary function $g_t$, the well-known analytical solution of equation (2.10) is:

$$\frac{c_t}{c_0} = \frac{1}{\alpha + (1-\alpha)e^{-G_t}} \tag{2.11}$$

Where $c_0$ denotes the initial level of RPCC at time $t = 0$, $\alpha \equiv c_0/C$, and $G_t = \int_0^t g_{t'} dt'$. From here, if $G_{t \to \infty} = +\infty$, this solution tends to $c_t = C$, which implies that eventually the dependency of the solution (2.11) on the initial condition will be lost and the value of RPCC will be completely determined by the parameter $C$. The beauty of the logistic model lies in its simplicity and the interpretability of its two parameters. In particular, when $c_t$ is much smaller than its long-term limit $C$, the intrinsic rate $g_t$ corresponds to the instantaneous rate of RPCC growth.

It is easy to see that if the intrinsic growth rate is assumed to be constant, $g_t = g$, these expressions yield the discount curve $y_t$ which is monotonically declining with time to $y_{t \to \infty} = \delta$:

$$y_t = \delta + \gamma g - \gamma t^{-1} \ln\{1 + \alpha[\exp(gt) - 1]\} \tag{2.12}$$

As expected, a declining with time relative growth rate of RPCC

$$c_t^{-1} \frac{dc_t}{dt} = \frac{g \exp(-gt)}{\alpha/(1-\alpha) + \exp(-gt)} \tag{2.13}$$

leads to a declining long-term tail of a social discount curve, Eq.(2.12). We would like to emphasize here that this effect is solely due to dynamic constrains imposed on the consumption growth process by environmental boundaries. Notably, Eq.(2.12) reduces to the classic Ramsey formula, $y_t = \delta + \gamma g$, for time-horizons much shorter than the characteristic time determined by the inflection point of the logistic curve, Eq.(2.11)

$$t_c = g^{-1} \ln \frac{1-\alpha}{\alpha} \tag{2.14}$$

Only in this limit the consumption growth rate depends linearly on the current level of consumption, the Verhulst equation (2.10) can be approximated by $dc_t/dt = g_t c_t$, and the negative third term in the RHS of Eq.(2.12) can be neglected!

### c. Uncertainty in future consumption growth rates. The cumulant expansion

Now let us consider a generally non-stationary stochastic nature of the consumption growth process. In our model the intrinsic rate $g_t$ is fluctuating in time due to random exogenous macroeconomic shocks, whereas $c_t$ is responding to these fluctuations. Our goal is to evaluate the expected discount rate under some *realistic* assumptions about the statistical characteristics of the stochastic process $g_t$ admitting tractable analytical results. The expected growth curve of log-consumption, from Eq.(2.11), can be written in the following form

$$\ln E[c_t / c_0] = \ln E[R(-1, G_t)] \tag{2.15}$$

where we introduce

$$R(\gamma, x) \equiv [\alpha + (1-\alpha)e^{-x}]^\gamma \tag{2.16}$$

Substitution of the solution (2.15) into Eq. (2.9) leads to

$$y_t = \delta - t^{-1} \ln E[R(\gamma, G_t)] \tag{2.17}$$

Notice that Eqs.(2.9) and (2.15) - (2.17) resemble many problems in natural sciences, where the effective long-term rate of growth (or decay) of an observable quantity can be time-dependent due to a stochastic process which modulates its instantaneous rate of growth (or decay). The key task in these problems reduces to averaging of a functional of a driving stochastic variable [7-10].

First, let us "de-trend" the non-stationary stochastic process $G_t$, by introducing

$$m_g(t) \equiv \frac{E[G_t]}{t} = \frac{1}{t}\int_0^t E[g_{t'}]dt' \tag{2.18}$$

so that it can be re-written as $G_t = m_g(t)t + \tilde{G}_t$, where

$$\tilde{G}_t = \int_0^t \tilde{g}_{t'}\, dt' \tag{2.19}$$

and $\tilde{g}_t \equiv g_t - E[g_t]$. Thus, $\tilde{G}_t$ and $\tilde{g}_t$ represent zero-mean fluctuations of $G_t$ and $g_t$, respectively. Finiteness of all moments of $\tilde{G}_t$ guarantees existence of the *converging* cumulant expansion [7-10]

$$\ln \chi(s,t) \equiv \ln E[\exp(s\tilde{G}_t)] = \sum_{n=2}^{\infty} \frac{s^n}{n!} Q_n(t) \tag{2.20}$$

Here, by definition, $\chi(s,t)$ is the moment-generating function and $Q_n(t)$ represents the *n*-th cumulant of $\tilde{G}_t$. To proceed further in the case of a rather general stochastic process $\tilde{g}_t$, which is not assumed to be either Normal, or i.i.d., or Markovian, or even stationary, we employ the approach that has been developed by Kubo [7]. It follows from Eqs. (2.19) and (2.20) that

$$Q_n(t) = \int_0^t dt_1 ... \int_0^t dt_n\, q_n(t_1,...,t_n), \quad n \geq 2 \tag{2.21}$$

Here $q_n(t_1, ... t_n)$ represent joint cumulants of random variables $\tilde{g}_{t_1}, ... \tilde{g}_{t_n}$. Every $n$-th cumulant $q_n(t_1, ... t_n)$ can be expressed via the first $n$ moments $m_n(t_1, ... t_n) \equiv E[\tilde{g}_{t_1} ... \tilde{g}_{t_n}]$. In particular, for central moments the first three generally non-zero cumulants are expressed as follows

$$q_2(t_1,t_2) = m_2(t_1,t_2),$$

$$q_3(t_1,t_2,t_3) = m_3(t_1,t_2,t_3)$$

$$q_4(t_1,t_2,t_3,t_4) = m_4(t_1,t_2,t_3,t_4) - m_2(t_1,t_2)m_2(t_3,t_4) - m_2(t_1,t_3)m_2(t_2,t_4) - m_2(t_1,t_4)m_2(t_2,t_3)$$

If $\chi(s,t)$ exists, expressions (2.20) and (2.21) appear to be as far as we can go, maintaining full generality.

Let us assume now that the average action of every single shock to $\tilde{g}_t$ during a characteristic time $\tau_g$ is small: $\rho_g \tau_g \ll 1$, where

$$\rho_g(t) \equiv \sqrt{q_2(t,t)} \tag{2.22}$$

$$\tau_g(t) \equiv \frac{1}{2q_2(t,t)} \int_{-\infty}^{\infty} q_2(t'-\frac{\tau}{2}, t'+\frac{\tau}{2}) d\tau \tag{2.23}$$

The motivation behind the explicit form of the definition (2.23) will become clear soon. For brevity, wherever possible we omit the explicit time argument of $m_g(t)$, $\rho_g(t)$, and $\tau_g(t)$.

For short times, $t \lesssim \tau_g$, from Eq. (2.21) we have $Q_n(t) \sim q_n(t,...,t)t^n$. Assuming that, while the parameter $\rho_g \tau_g$ tends to zero, all ratios $m_n(t,...t)/\rho_g^n$ remain finite, i.e. the "shape" of the distribution of $\tilde{g}_t$ is fixed, we also have $q_n(t,...t) \sim m_n(t,...t) \sim \rho_g^n$. Therefore,

$$Q_n(t) \sim (\rho_g t)^n \lesssim (\rho_g \tau_g)^n \tag{2.24}$$

In the opposite case, $t \gg \tau_g$, since the dominant contribution to the integral in Eq. (2.21) is produced in the vicinity of the diagonal $t_1 = t_2 ... = t_n$ with the characteristic width $\tau_g$, we arrive to the following estimate

$$Q_n(t) \sim \int_0^t q_n(t,...,t)\tau_g^{n-1} dt \sim \frac{t}{\tau_g}(\rho_g \tau_g)^n \tag{2.25}$$

Thus, when the key parameter $(s\rho_g \tau_g)^2$ is small, the contribution of higher cumulants $Q_n(t)$, $n > 2$, into the expansion Eq. (2.20), for *any* fixed "shape" of the distribution of $\tilde{g}_t$ is proportionally small for all times, which guarantees finiteness of all moments of $\tilde{G}_t$ and, hence, existence of the converging cumulant expansion (2.20). This crucial point leads to the following two important results. First, in the limitless economy, $\alpha = 0$, Eqs. (2.17) and (2.20) yield

$$y_t = \delta + \gamma m_g - t^{-1}\ln\chi(-\gamma, t) \tag{2.26}$$

and hence for $(\gamma \rho_g \tau_g)^2 \ll 1$ we immediately obtain

$$y_t \simeq \delta + \gamma m_g - t^{-1}\gamma^2 \frac{Q_2(t)}{2} \tag{2.27}$$

Secondly, at times $t \gg \tau_g$ the probability density function of $G_t$ has a sharp maximum around the mean value $m_g t$, with the variance $D_g(t)t$, where

$$D_g(t) \equiv \frac{Q_2(t)}{t} \simeq \frac{2}{t}\int_0^t \rho_g^2(t')\tau_g(t')dt' \tag{2.28}$$

As a result, in the bounded economy, the expectation of the functional $R(G_t)$ in Eq. (2.17) admits the approximate analytical estimation. Indeed, from Eq. (2.20) we have the characteristic function of $\tilde{G}_t$

$$\varphi(s,t) \equiv \chi(is,t) = \exp\left[\sum_{n=2}^{\infty} \frac{(is)^n}{n!} Q_n(t)\right] \tag{2.29}$$

Hence, the probability density function (PDF) of $G_t = m_g t + \tilde{G}_t$ is given by the inverse Fourier transformation

$$f(x,t) = \frac{1}{2\pi}\int_{-\infty}^{\infty} \exp\left[-is(x - m_g t) - s^2\frac{D_g t}{2} + \sum_{n=3}^{\infty} \frac{(is)^n}{n!} Q_n(t)\right] ds \tag{2.30}$$

For brevity we omit hereafter the explicit time argument of $D_g(t)$. Changing here the integration variable to $\bar{s} = s\sqrt{D_g t}$, we obtain:

$$f(x,t) = \frac{1}{\sqrt{D_g t}} \bar{f}\left(\frac{x - m_g t}{\sqrt{D_g t}}, t\right), \tag{2.31}$$

where

$$\bar{f}(x,t) = \frac{1}{2\pi} \int_{-\infty}^{\infty} \exp\left[-i\bar{s}x - \frac{\bar{s}^2}{2} + \sum_{n=3}^{\infty} \frac{(i\bar{s})^n}{n!} \frac{Q_n(t)}{(D_g t)^{\frac{n}{2}}}\right] d\bar{s} \qquad (2.32)$$

Taking into account the estimate (2.25), we see that, when $t/\tau_g \to \infty$, the sum of higher moments in Eq.(2.32) vanishes, so that $\bar{f}(x,t)$ approaches the normalized Gaussian bell. Thus, as expected in this limit, the distribution of $G_t$ approaches the normal distribution with the mean $m_g t$ and variance $D_g t$. What is important for the following however is only the fact that, when $t \gg \tau_g$, the PDF $f(x,t)$ of $G_t$ has a sharp maximum around the mean value $m_g t$ and the variance of this variable is $D_g t$. Therefore, the integral

$$E[R(\gamma, G_t)] = \int_{-\infty}^{\infty} R(\gamma, x)\bar{f}\left(-\frac{(x - m_g t)^2}{2D_g t}, t\right) \qquad (2.33)$$

may be approximated by the first two non-zero members of the "saddle point" asymptotic expansion like this:

$$E[R(\gamma, G_t)] \simeq R(\gamma, m_g t) + R''(\gamma, m_g t)\frac{D_g t}{2} \qquad (2.34)$$

From here, with the same accuracy,

$$\ln E[R(\gamma, G_t)] \simeq \ln R(\gamma, m_g t) + \frac{R''(\gamma, m_g t)}{R(\gamma, m_g t)} \frac{D_g t}{2} \qquad (2.35)$$

Substituting Eq.(2.35) into Eqs.(2.15) and (2.17) we obtain the following results

$$\ln E[c_t / c_0] \simeq \ln R(-1, m_g t) + \frac{R''(-1, m_g t)}{R(-1, m_g t)} \frac{D_g t}{2} \qquad (2.36a)$$

$$y_t \simeq \delta - \frac{\ln R(\gamma, m_g t)}{t} - \frac{R''(\gamma, m_g t)}{R(\gamma, m_g t)} \frac{D_g}{2} \qquad (2.36b)$$

where $R''(\gamma, x)$ denotes the second derivative of $R(\gamma, x)$ by $x$. These estimations comprise the main analytical results of this paper. Notably, formulas (2.36) are free from the model of a stochastic process and from the specific model of a negative feedback loop leading to declining with time relative rates of consumption growth.

## III. Analysis of limiting cases.

Let us explore explicit analytical results coming from Eqs.(2.27) and (2.36) in some limiting cases. Obviously, in the absence of fluctuations, $D_g = 0$, Eqs.(2.36) reduce to Eqs.(2.11) and (2.12), derived in the dynamic limit with a time invariant RPCC growth rate $m_g(t) = g$. From the definition of $R(\gamma, x)$ given by Eq. (2.16) we have

$$\frac{R''(\gamma, x)}{R(\gamma, x)} = \gamma^2 (1-\alpha) \frac{1-\alpha+\alpha e^x / \gamma}{(1-\alpha+\alpha e^x)^2} \tag{3.1}$$

Substitution of this expression into Eqs. (2.36) yields the following asymptotic results in the limiting cases of short and long times, relative to the characteristic time determined by Eq. (2.14) (with $g$ identified as the average $m_g$):

$$y_t \simeq \delta + \gamma m_g \left[ 1 - \frac{\ln(1-\alpha+\alpha e^{m_g t})}{m_g t} \right] - \gamma^2 \frac{D_g}{2}, \qquad t_c - t \gg m_g^{-1} \tag{3.2a}$$

$$y_t \simeq \delta - \frac{\gamma \ln \alpha}{t} - \gamma \frac{1-\alpha}{2\alpha} D_g e^{-m_g t}, \qquad t - t_c \gg m_g^{-1} \tag{3.2b}$$

and

$$\ln E[c_t / c_0] \simeq m_g t \left[ 1 - \frac{\ln(1-\alpha+\alpha e^{m_g t})}{m_g t} \right] + \frac{D_g}{2} t, \qquad t_c - t \gg m_g^{-1} \tag{3.3a}$$

$$\ln E[c_t / c_0] \simeq -\ln \alpha - \frac{D_g t}{2} \frac{1-\alpha}{\alpha} e^{-m_g t}, \qquad t - t_c \gg m_g^{-1} \tag{3.3b}$$

Formulas (3.4) permit direct comparison with empirical data [6], which we conduct in the next section.

The product $\gamma m_g$ of the degree of concavity of the power utility function ($\gamma > 0$) and the positive expected intrinsic rate of consumption growth $m_g$ in Eq.(3.2a) increases the social discount rate of benefits deferred to the distant future. Since the key parameter $\alpha < 1$, the second term in the RHS of Eq.(3.2b) is also positive. However, its role is hyperbolically *decreasing* with time. The negative third terms in the RHS of Eqs. (3.2a) and (3.2b) are interpreted as the 'precautionary effect'. These terms quantify the Jensen inequality in the bounded economy. The relative accuracy of these terms is determined by the two small parameters: $(\gamma \rho_g \tau_g)^2 \ll 1$ and $\tau_g/t \ll 1$. In the economy with a constrained consumption growth, in the long-term limit the magnitude of the 'precautionary' term is exponentially decreasing with time. At an intuitive level, this can be explained by a simple argument - preventing the near-certainty of collision with an asteroid in the next million years is much less of concern of a government today than a relatively unlikely event of a crash in the next hundred years. Remarkably, due to convexity of the *growing* consumption at short times, the Jensen's correction term in Eq.(3.3a) is positive, leading to higher growth rates than estimates based on the historical average. On the other hand, in the long-term limit, when a consumption growth process will be inevitably slowing down, the RPCC $c_t$ will become a *concave* function of time and the Jensen's correction term in Eq.(3.3b) turns out to be negative.

Due to the definition (2.28) and that

$$\lim_{t \to 0}\left[1+\frac{\ln(1-\alpha+\alpha e^{m_g t})}{m_g t}\right]=1+\alpha$$

it is easy to see that in the unbounded economy, $\alpha=0$, the short-time asymptotic (3.2a) and (3.3a) give

$$y_t \simeq \delta + \gamma m_g - \gamma^2 t^{-1}\int_0^t \rho_g^2(t')\tau_g(t')\,dt' \tag{3.4a}$$

$$\ln E[c_t/c_0] \simeq m_g t + \int_0^t \rho_g^2(t')\tau_g(t')\,dt' \tag{3.4b}$$

with the relative accuracy of the order $(\gamma \rho_g \tau_g)^2 \ll 1$. Clearly, the same results proceed directly from Eq. (2.27), when $t \gg \tau_g$. These results are further simplified for a stationary stochastic process $\tilde{g}_t$, when $D_g \simeq 2\rho_g^2\tau_g$. In this case, the expected social discount rate reaches its time invariant asymptotic form

$$y_t \simeq \delta + \gamma m_g - \gamma^2 \rho_g^2 \tau_g \tag{3.5a}$$

whereas the expected log-consumption is linearly growing with time:

$$\ln E[c_t/c_0] \simeq (m_g + \rho_g^2 \tau_g)t \tag{3.5b}$$

It is instructive to see the outcome of Eq. (2.27) without the assumption $t \gg \tau_g$. In this case, we have to introduce a specific model of the stochastic consumption growth process. For example, the Markov (*M*) model with an exponentially decreasing auto-covariance, $q_2(t_1, t_2) = \rho_g^2 \exp(-\tau/\tau_g)$, where $\tau = |t_1 - t_2|$, yields

$$Q_2(t) = 2\rho_g^2 [\tau_g t - \tau_g^2 (1 - e^{-t/\tau_g})] \tag{3.6}$$

Hence, in this model

$$y_t^{(M)} = y_\infty^{(M)} + \gamma^2 \rho_g^2 \tau_g^2 t^{-1} (1 - e^{-t/\tau_g}) \tag{3.7}$$

where

$$y_\infty^{(M)} = \delta + \gamma m_g - \gamma^2 \rho_g^2 \tau_g \tag{3.8}$$

consistently with Eq.(3.5a). Within the domain of validity of the stochastic perturbation theory, $(\gamma \rho_g \tau_g)^2 \ll 1$, the time-invariant $y_\infty^{(M)}$ represents the lowest possible value of the long-term risk-free interest rate in the boundless economy with a linear consumption dynamic. This regime is inevitable on a coarse grained time scale, when a dependency of the result on the path of $g_t$ between time 0 and *t* is lost. However, even in this limiting case, the value of a long-term risk-free interest rate depends on the time scale $\tau_g$ of a serial correlation in spot rates of a log-consumption growth process. Formula (3.7) tells us that in the boundless economy the decline of $y_t$ could be significant only at times comparable with the autocorrelation time $\tau_g$.

Alternatively, one could derive model-dependent results by postulating, e.g., that fluctuations of $\tilde{g}_t$ follow the Ornstein-Uhlenbeck (OU) stochastic process [29]. In the long-term this model leads to the Gaussian stationary distribution of $\tilde{g}_t$. In this model all cumulants of higher than the second order vanish identically. Consequently, within the OU model the exact expression for the long-term tail of the risk-free interest rate is structurally similar to that of Eq.(3.7):

$$y_t^{(OU)} = y_\infty^{(OU)} + \frac{\gamma^2 \sigma_g^2}{2\alpha_g^3 t} (1 - e^{-\alpha_g t}) \tag{3.9}$$

$$y_\infty^{(OU)} = \delta + \gamma m_g - \frac{\gamma^2 \sigma_g^2}{2\alpha_g^2} \tag{3.10}$$

Here $\alpha_g$ characterizes the speed of reversion to the normal level of the average rate $m_g$, whereas $0.5\sigma_g^2$ is the coefficient of diffusion in the space of spot rates, measured in 1/time$^3$. This model is conceptually analogous to the Vasicek model of the term structure of interest rates [30]. Expressions (3.9) and (3.10), with an obvious redefinition of parameters, correspond to the long-term asymptotic of the original Vasicek result. They, once again, show that account of serial correlation in an instantaneous log-consumption growth rates leads to a declining schedule of long-term risk-free interest rates at times comparable with $\tau_g = 1/\alpha_g$, see also the comprehensive review of the memory related effects in Ref.[2]. Expressions (3.7) - (3.10) were first derived by one of the authors [31]. However, the exposition of the method and main results in this paper were contaminated with a number of errors in definitions of the key variables.

## IV. Comparison with empirical data.

The benchmark model of consumption-based asset pricing and cost-benefit analysis of public projects [2 4] assumes that $\ln c_t$ follows the arithmetic Brownian motion with a constant effective drift $m_g$ and time-invariant annualized volatility $\sigma$, leading to the time-invariant discount rate, $y = \delta + \gamma m_g - 0.5\gamma^2\sigma^2$. We begin this section with an assessment of consistency of this model with data [6]. The upper panel of Fig.2 shows that the evolution of the real (corrected for inflation) U.S. per capita log-consumption between 1889 and 2009 can be very well approximated by the simple linear growth model with a positive trend of 2.0% per year. Note, however, that during the Great Depression the real U.S. RPCC decreases by a cumulative 26% over the five year period between 1930 and 1934. This drop was preceded by a five-year increase by 12% during 1926 -1930 and a subsequent growth by 21% between 1934 and 1938. The dataset was collected by Robert Shiller [6], where readers can find details of the data collection and valuation methodology. The irregular behavior of annual increments of real per capita log-consumption growth, $\ln c_t / c_{t-1}$, is shown in the lower panel of Fig.2. Although, the visual inspection of the time series clearly points to a big difference between the pre- and post-war regimes, both augmented Dickey–Fuller and Kwiatkowski–Phillips–Schmidt–Shin tests support the trend-stationarity of the empirical time series between 1889 and 2009. This discrepancy can be related to a small sample size.

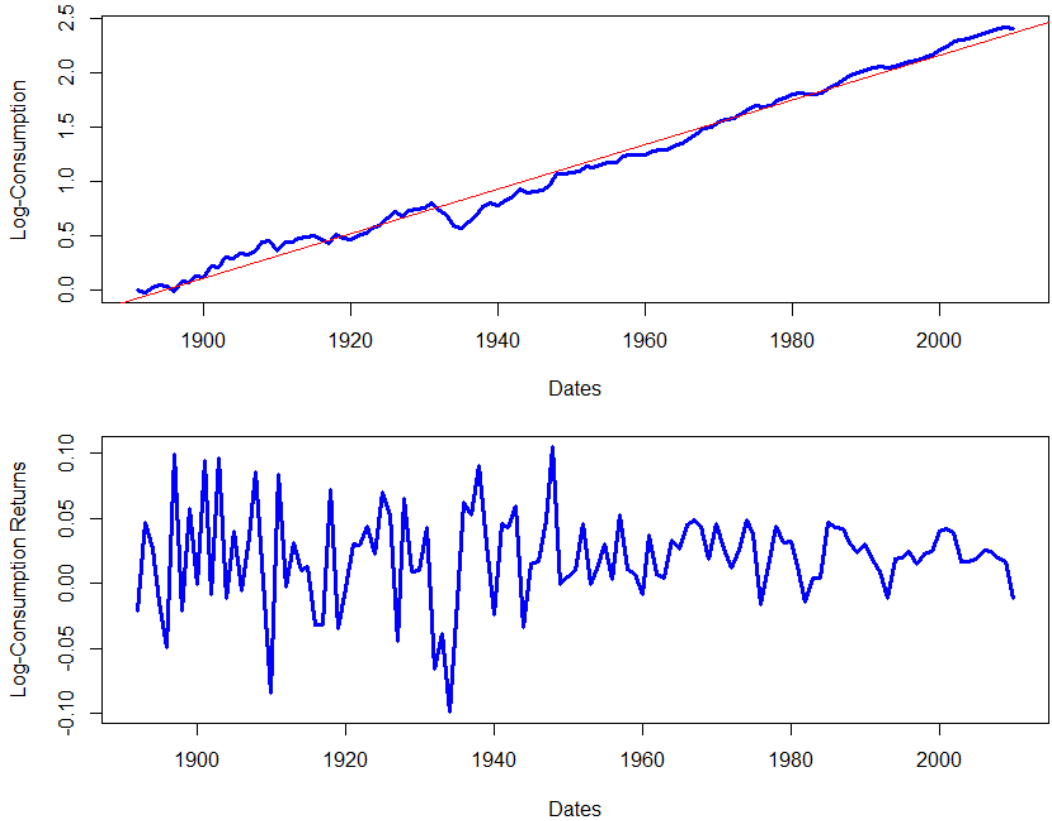

Fig.2. Evolution of the real U.S. per capita log-consumption, $\ln c_t / c_0$, (upper panel) and its annual increments, $\ln c_t / c_{t-1}$ (lower panel) between 1889 and 2009. The straight line in the upper panel represents the fitted linear model; intercept = 7.82, slope = 0.021/year, R-squared = 0.99, $p < 2.2\text{e-}16$.

The first four moments of the distribution of $\ln c_t / c_{t-1}$ are as follows: the mean, $m_g = 2.0\% / year$, the variance, $\rho_g^2 = 0.00123 / year^2$, the skewness = -0.38, the excess kurtosis = 1.14. Similar values were reported in Refs.[32,33]. The estimated values of higher moments depend on the software package used for valuation and are not definitive, which is typical for fat-tailed distributions and small sample size. Nevertheless, the excess kurtosis is convincingly far from zero, pointing to a non-Gaussian character of the distribution. This conclusion is also supported by visual inspection of the empirical density distribution of $\ln c_t / c_{t-1}$ and the relevant Q-Q plot (not shown here) as well as Jarque-Bera, Shapiro-Wilk, and Wilcoxon tests, which strongly reject the hypothesis of normality of the distribution of increments $\ln c_t / c_{t-1}$.

Moreover, it is evident from Figure 3 that the variance of $\ln c_t / c_{t-1}$, calculated with a yearly rolling window of 20 years, is substantially varying in time. This finding, which does not depend on the size of the rolling window (up to 50 years) imply that the underlining process is non-stationary; increments of log-consumption growth are not *identically* distributed over time, which is also visually clear from the lower panel of Fig.2. Thus, the second assumption of the benchmark model - the increments of log-consumption growth are identically distributed - is not supported by data.

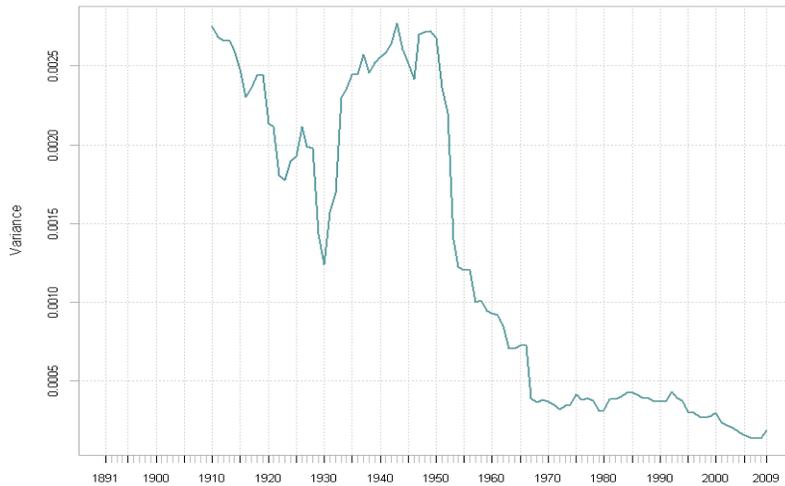

Figure 3. The temporal behavior of the variance of annual increments of real per capita log-consumption growth calculated with a rolling window of 20 years.

Appearance of small, yet statistically significant at 5% confidence interval, picks at lags 9 and 10 in correlograms of $\ln c_t / c_{t-1}$ (not shown here) suggests a rather long serial correlation time. On the other hand, the Box-Pierce test does support the null hypothesis of independently distributed log-consumption returns, $p$-value = 0.43. The small sample size and the non-Gaussian character of the distribution under study make interpretation of these results inconclusive, see, e.g., Ref. [34]. Furthermore, the analysis of Beeler and Campbell [35] of the U.S. RPCC does not support an existence of relatively long memory in fluctuations of the log-consumption growth rates. Conversely, Cochrane [36] and Cogley [37] have shown that the presence of serial correlation in growth rates of log-consumption is supported by data in most countries. The observed strict exponential growth of U.S. RPCC can be described by Eq.(3.6b), which implies the Markovian character of the evolution of $c_t$, i.e., a short memory between shocks in $g_t$.

Interestingly, the Box-Pierce test and correlograms of squared as well as absolute values of $\ln c_t / c_{t-1}$ (not shown here), which are often used as proxies for a variance of the relevant stochastic process, do not support manifestation of serial correlations. Hence, the time series of annual increments of per capita log-consumption do not exhibit a "volatility clustering" that is typical to log-returns of stocks. This observation contradicts the key assumption of the model with persistent variations in consumption volatility, see Ref.[13] and references therein.

To summarize, evidently both key assumptions of the benchmark model: i) the Gaussian distribution of $\ln c_t / c_{t-1}$ and ii) the Brownian diffusion model of a log-consumption growth process are not valid. Accordingly, the generalization of the Ramsey formula derived within the traditional model of arithmetic Brownian motion and its extension to fat-tailed distributions of increments of log-consumption growth made within the i.i.d. model of the consumption growth process are not supported by empirical observations of the U.S. RPCC growth between 1889 and 2009.

The formalism that has been developed in Section 2 is not restricted by these modeling assumptions. It is important to stress here that empirically the variance of annual increments of log-consumption is rather small, $\rho_g^2 = 0.00123 / year^2$. This is not surprising for this process is not directly dependent on typical behavioral factors driving large fluctuations of stock-market returns, e.g., herding, overreaction, etc. Therefore, the condition of validity of the theoretical framework developed in Section 2, $(\rho_g \tau_g)^2 \ll 1$, is satisfied for long autocorrelation times $\tau_g \leq 10$ years! Moreover, if $\tau_g \leq 5$ years, even the more stringent validity condition, $(\gamma \rho_g \tau_g)^2 \ll 1$, required by the fast convergence of Kubo cumulant expansion for the discount rate, is satisfied for the widely accepted value $\gamma = 2$ of the parameter of an intergenerational inequality aversion, $(\gamma \rho_g \tau_g)^2 \leq 0.123$.

To see the practical application of the valuation based on the developed theory, let us go back in time to 1949 and try to forecast the consumption level in 2009. In the unbounded economy, if we ignore the positive Jensen correction term in the RHS of Eq.(3.5b), the forecast is trivial, but wrong, see Fig.4. Indeed, using information available at the end of 1949, $m_{g,1949} = 1.82\%$ per year, $c_{1949} = \$8,045.5$ (in 2005 U.S. dollars [6]), the estimate would be $c_{2009} = \$23,932.2$, which is 21.6% lower than the actual consumption level of \$30,509 in 2009. The size of the error is substantial and would be much larger for longer time horizons. On the other hand, if we employ Eq.(3.5b) with the estimated mean and variance of log-consumption return for the period between the end of 1890 and the end of 1949, $m_g = 1.82\%$ per year and $\rho_g^2 = 0.0022/\text{year}^2$, we obtain a very good fit to observations with the reasonable autocorrelation time: $\tau_g = 1.87$ years. Now let us plug these parameters into Eq.(3.6a) and use the plausible values of $\gamma = 2$ and the ethical value of the 'impatience' parameter $\delta = 0$, see Refs.[1 2]. The negative Jensen correction term reduces the estimated discount rate by 1.65%. This leads to the reasonable estimate of the expected value of a long-term social discount rate, $y_{2009} = 1.99\%$ per year, which should be used for cost-benefit analysis of public projects with benefits deferred for 2009, based on data available to statistician at the end of 1949.

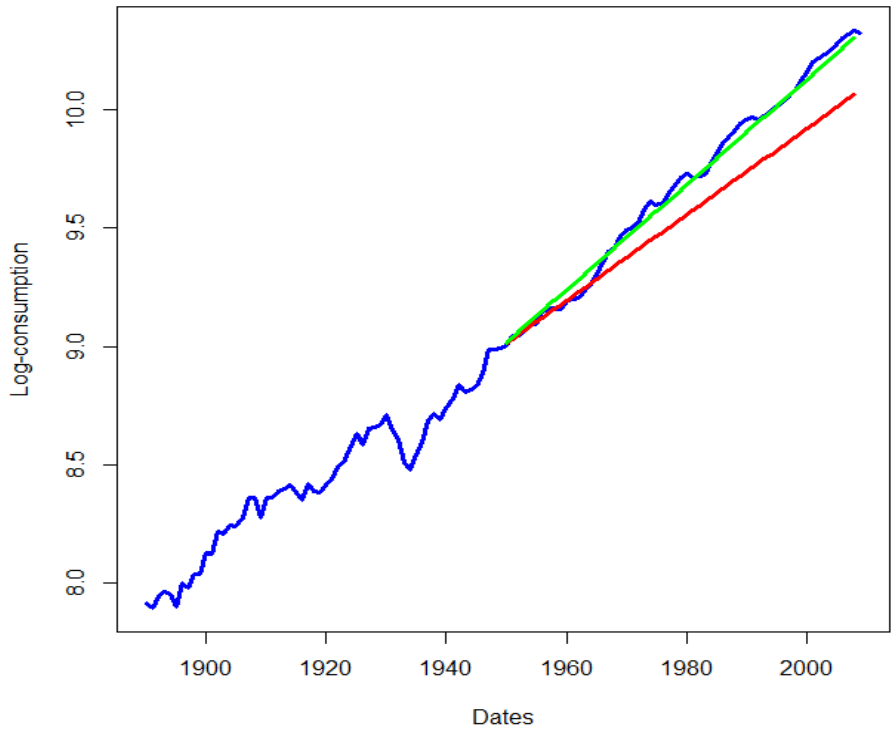

Figure 4. (Color online). Forecast of the U.S. real per capita log-consumption growth based on data available at the end of 1949. Green and red straight lines represent forecasts based on Eq.(3.6b) with and without stochastic term, respectively; $c_{1949} = \$8,045.5$, estimated $m_{g,1949} = 1.82\%$ per year, $\rho_g^2 = 0.0022/\text{year}^2$, and fitted $\tau_g = 1.87$ years.

The observed strict exponential growth of RPCC in the USA implies that we are still far from the inflection point in its evolution and conventional assumption of a boundless economy is still satisfied. Plugging the estimated average rate of RPCC growth $g = 2.0\%$ per year into Eq.(2.14) we obtain the rough estimate of the inflection point $t_c \simeq 50\ln(C/c_0)$ years. Inspection of the logistic function (2.11) shows that deviation from the strict exponential growth of RPCC is detectable for relatively big values of $\alpha > 0.1$. Therefore, the lower boundary of $t_c$ is at least 100 years from 2009. Nonetheless, logarithm is the notoriously slow-growing function at large values of its argument. Hence, even if $\alpha \simeq 10^{-4} \div 10^{-2}$, the characteristic time, $t_c \simeq 200 \div 400$ years, is not that far from now. Figure 5 shows plots of the expected long-term structure of social discount rates in the bounded economy forecasted by Eqs.(2.36b) and (3.1) for different values of $\alpha = \{0.1, 0.03, 0.01\}$. To quantify this behavior, we choose $\gamma = 2$ and the ethical value of the parameter $\delta = 0$. The values of the first and second moments of the RPCC growth process were estimated from the observed time series between the end of 1979 and the end of 2009: $m_g = 2.0\%$ per year and $D_g = 0.0012$ per year. Plots on Fig.5 show that when $c_0 \ll C$, the 'precautionary' effect induced by uncertainty in the future level of RPCC is the dominant mechanism of declining discount rates. The role of this effect is reflected by lower lines in 'doublets' corresponding to the same $\alpha$ on Fig.5. However, with growth of time, the dynamic effect related to a negative feedback mechanism which constrains the RPCC growth process, becomes dominant. The role of this effect is reflected by upper lines in 'doublets' corresponding to the same $\alpha$ on Fig.5. It is easy to see from these plots that in the not that 'distant future', the discounting policy satisfying Ramsey's optimization of social welfare functional would dramatically change. Future generations, facing a slowing down growth of RPCC due to biophysical boundaries of our planet, will almost surely adjust their social discounting policy towards much higher value of long-term future benefits than it is currently accepted. Note that the inflection point in the logistic model corresponds to $t_c = 0$ for $\alpha = 0.5$. Far beyond this point Eq.(3.3a) forecasts the hyperbolic decrease of social discount rates with time.

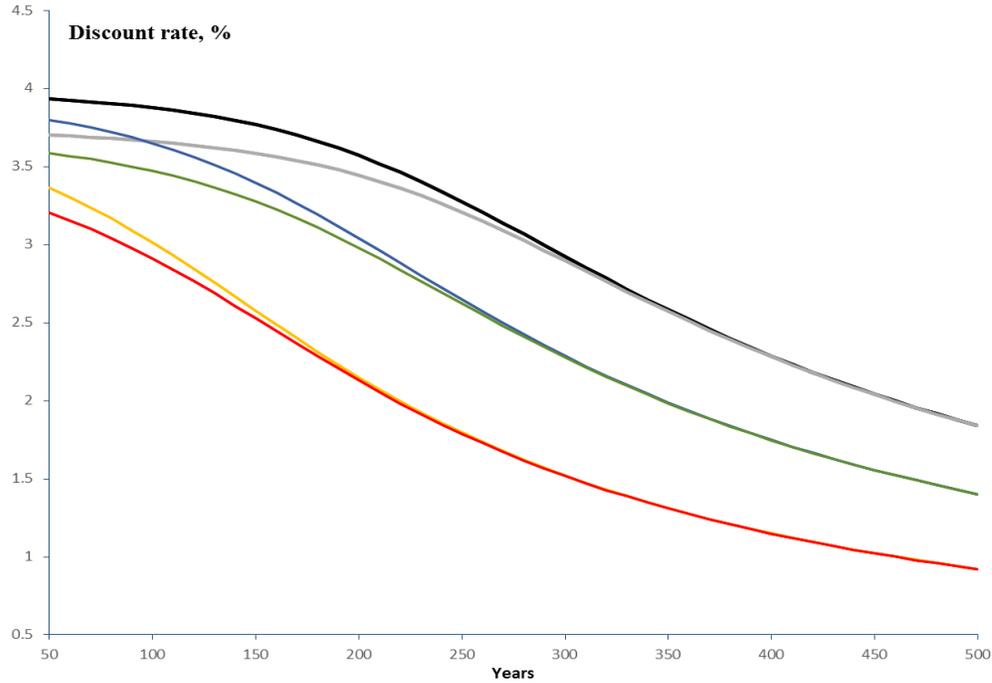

Figure 5. The expected long-term structure of social discount rates forecasted by Eqs.(2.36b) and (3.1), $\delta = 0$, $\gamma = 2$, $m_g = 2.0\%$, and $\alpha = \{0.1, 0.03, 0.01\}$ from bottom to top, respectively. The lower lines in each of the 'doublets' correspond to $D_g = 0.0012$ per year, whereas the upper lines correspond to pure dynamic discounting, $D_g = 0$. See text for details.

## V.     Summary and discussion.

Due to fundamental constrains imposed by finite resources of our planet, the exponential growth of RPCC is not sustainable indefinitely into the future. Therefore, in the distant future its growth rates will imminently start to decline with time. Our study has applied this basic insight to valuation of long-term social discount rates. Combining the proposed logistic model of consumption growth with the Ramsey optimal growth framework and power utility, we demonstrate that a negative feedback loop, which prevents an unlimited growth of consumption, creates a strong *dynamic* effect leading to declining long-term tail of social discount curve. This conclusion is very important for the climate finance with benefits on current investments deferred to centuries from now.

Clearly, the growth rate of consumption is irregularly evolving in time. Comparison with empirical data shows that traditional stochastic models are not fitting observations. However, the product of an average strength of fluctuations in the growth rate and its autocorrelation time is very small. This empirical fact allows for truncation of the Kubo cumulant expansion at the second term, leading to remarkably simple expressions for the term-structure of *expected* consumption growth and associated discount rates in a limitless as well as in a *bounded* economy. The logistic model yields the characteristic point in time $t_c$ when the RPCC growth starts to slow down. However, the observed U.S. RPCC between 1889 and 2009 is growing strictly exponentially. Therefore, we were able to estimate only the lower boundary of $t_c$ as 100 years from 2009. Beyond this time horizon the dynamic effect related to planetary resource constrains will become a dominant mechanism responsible for slowing down of a consumption growth and, hence, the declining long-term tail of a social discount curve.

Obviously, the proposed model ignores the possibility of the carrying capacity of RPCC to decrease with time due to the population growth as well as a stronger negative influence on consumption, which real world system may exhibit. Moreover, in the context of the climate finance, the carrying capacity of RPCC is itself an uncertain function of rising temperature due to greenhouse gas emissions from human activities. The model introduced in this paper disregards geographic heterogeneity of RPCC growth rates, intergenerational fear and 'impatient' factors. A negative feedback loop leading to declining growth rates of RPCC might be stronger than it is assumed by the logistic model.

On the other hand, the theoretical framework that has been developed in this paper is free from conventional assumptions of stationarity, normality, and i.i.d. fluctuations of log-consumption growth that are not supported by empirical data. It explicitly takes into account constrains imposed on the consumption growth process by planetary boundaries, which become increasingly important for valuations of social discount rates as the time horizon lengthens. It is important to emphasize here that with the relevant modification of the function $R(\gamma, x)$, our key results, Eqs.(2.36), are not limited by the logistic model of RPCC growth. Moreover, the formalism developed in Section 2 is not restricted by the power utility. It allows for a straightforward extension to diverse utility functions. We demonstrate that the assumed smallness of an average action of stochastic shocks to RPCC growth rates is empirically sound.

Although, unforeseeable future economic and non-economic events, may lead to catastrophic shocks to consumption growth rates with high persistence which could invalidate the applicability of the truncated cumulant approach employed in this paper. Nonetheless, we believe that our study illuminates the crucial role that account of planetary boundaries should play in shaping a rigorous long-term social discounting policy.

Although this is not the focus of this paper, the obvious redefinition of the key parameters of the model, allows for a straightforward application of the obtained formal results to estimations of an expected long-term population growth curve in stochastic environments, which is relevant in every field of biology and demography that is concerned with nonlinear growth processes. We are planning to explore this topic in a near future.

**Acknowledgement**

Authors are grateful to Professor I. Tkachenko, Dr. N. Shokhirev, and Professor J. Chen for helpful discussions on a range of problems dealt with in this paper. Research of Y.K. was supported by S&P Global Market Intelligence. The views expressed in this paper are those of the authors, and do not necessary represent the views of PTC and S&P Global Market Intelligence.

---